\documentclass[letter]{aa}  
\usepackage{graphicx}
\usepackage{txfonts}
\usepackage{lipsum}
\usepackage{subcaption}         
\usepackage{lscape}
\usepackage{placeins}
\usepackage{float}  
\usepackage[]{hyperref}
\providecommand{\msun}{\ensuremath{\mathrm{M}_\odot}}
\defcitealias{kumar24}{K24}

\begin{document}

   \title{The tightest binary post-AGB star}

   \author{R. Salinas\inst{1}\corrauth{rsalinas@camk.edu.pl} 
          \and C. Salgado\inst{2}\email{c.salgadoescalona@uandresbello.edu}
          \and N. Rodríguez-Segovia\inst{3}\email{njrsegovia@kasi.re.kr}
          \and C. Sáez-Carvajal\inst{4}\email{saezc.cata@gmail.com}
          \and V. Kalari\inst{5}\email{venu.kalari@noirlab.edu}
          \and M. Catelan\inst{6}\email{mcatelan@astro.puc.cl} 
          \and S. Kimeswenger\inst{7}\email{Stefan.Kimeswenger@uibk.ac.at}
          \and F. Marcadon\inst{8}\email{frederic.marcadon@gmail.com}
          \and P. Wielg\'orski\inst{1}\email{pwielgor@camk.edu.pl}  
          \and A. Escorza\inst{9,10}\email{ana.escorza.santos@iac.es}
          }
   \institute{Nicolaus Copernicus Astronomical Center, Polish Academy of Sciences, Bartycka 18,00-716 Warsaw, Poland                 
              \and 
              Universidad Andrés Bello, Facultad de Ciencias Exactas, Departamento de Física y Astronomía - Instituto de Astrofísica, Autopista Concepción-Talcahuano 7100, Talcahuano, Chile            
              \and 
              Korea Astronomy and Space Science Institute, 776 Daedeok-daero, Yuseong-gu, Daejeon 34055, Republic of Korea
              \and
              Instituto de Física y Astronomía, Universidad de Valparaíso, Av. Gran Bretaña 1111, 5030 Casilla, Valparaíso, Chile
              \and
              Gemini Observatory/NSF's NOIRLab, Casilla 603, La Serena, Chile
              \and 
              Instituto de Astrofísica, Pontificia Universidad Católica de Chile, Av. Vicuña Mackenna 4860, 7820436 Macul, Santiago, Chile
              \and 
              Universit{\"a}t Innsbruck, Institut f{\"u}r Astro- und Teilchenphysik,  Technikerstr. 25\/8, 6020 Innsbruck, Austria
              \and
              Private Researcher, Orléans, France
              \and
              Instituto de Astrofísica de Canarias, C. Vía Láctea, s/n, 38205, La Laguna, Santa Cruz de Tenerife, Spain
              \and
              Universidad de La Laguna, Dpto. Astrofísica, Av. Astrofísico Francisco Sánchez, 38206 La Laguna, Santa Cruz de Tenerife, Spain
              }
   \date{}

  \abstract
   {}
   {A new post-asymptotic giant branch (pAGB) star has recently been reported in the Galactic globular cluster E3. Only a handful of these systems are known. Based on two radial velocities it has been suggested that this pAGB is in a binary system. This work seeks for confirmation of this binarity and a more complete description of the system.} 
   {Time-series spectroscopy of the system was acquired with Gemini/GMOS together with a Gemini/GHOST high-resolution spectrum from which stellar parameters and chemical abundances were measured via a non-local thermodynamic equilibrium (NLTE) analysis. A reconstruction of its possible formation channels was done with the fast binary population synthesis (BPS) code \textsc{compas.}}
   {The system has an orbital period of 34.04 $\pm$ 0.05 days, making it the first confirmed binary pAGB in a Galactic globular cluster, and the one with the shortest period among all known binary pAGB stars. Unlike the majority of such systems, the chemical abundances of the pAGB star show no sign of depletion of refractory elements, consistent with the absence of a circumbinary disc. This absence may be connected to the complex formation process suggested by BPS models, where the pAGB star was originally the least massive star in the system, and the most massive component evolved into a present-day likely helium white dwarf.}
   {}


   \maketitle
   \nolinenumbers
   
\section{Introduction}\label{sec:intro}

Post-asymptotic giant branch stars (pAGBs) comprise a fast phase in the late evolution of low and intermediate-mass stars after the strong mass loss of the AGB has concluded, when the expelled material dissipates, and the leftover core increases its temperature while maintaining a close to constant luminosity \citep[e.g.][]{vanwinckel03}. This scenario can be further complicated in binary systems, with the formation of circumbinary discs impacting the processes of mass loss and dust formation \citep[e.g.][]{vanwinckel18}. 

Galactic globular clusters (GCs) host only about a couple of dozens pAGBs \citep[e.g.][]{dixon24}, with none of them belonging to a binary system. Recently, \citet[][hereafter K24]{kumar24} presented the discovery of a hot pAGB in the low-mass GC E3, based on observations with the Ultraviolet Imaging Telescope \citep[UVIT,][]{tandon17} and \textit{Gaia} astrometry, also proposing it as a member of a binary system but based solely on the discrepancy of two radial velocity (RV) measurements.

In this Letter we present 18 RV measurements of this pAGB with the Gemimi-S/GMOS, Gemini-S/GHOST and JK15/BESO spectrographs, taken over a timespan of 772 days (observation details are in Appendix \ref{appendix:data}), confirming its binarity and accurately determining its orbital parameters. The orbital modelling is presented in Sect.\ref{sec:rvs}. Stellar parameters and chemical abundances based on a high-resolution GHOST spectrum are given in Sect. \ref{sec:abundances}, followed by the results of binary synthesis in Sect. \ref{sec:synthesis}, while a discussion of the results is presented in Sect. \ref{sec:concl}.

\section{Orbit modelling}\label{sec:rvs}

\subsection{Orbital modelling and orbital parameters}

RVs were modelled with the Monte Carlo orbit sampler \textsc{TheJoker} \citep{thejoker}, using priors based on the loosely constrained parameters given by \citetalias{kumar24}.

\begin{figure}[t]
   \centering
   \includegraphics[scale=0.6]{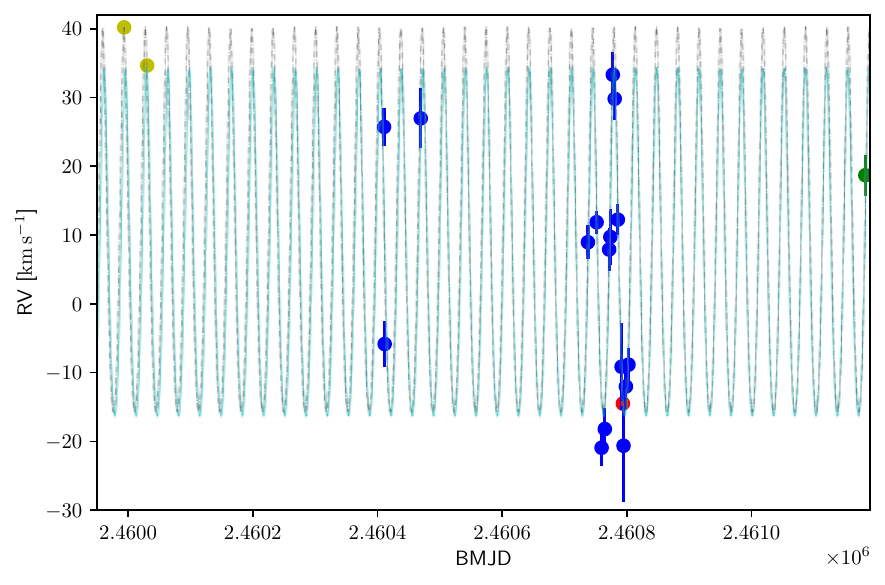}
      \caption{GMOS (blue circles), GHOST (red) and BESO (green) RVs and their best-fitting Keplerian model by \textsc{TheJoker} in cyan lines. The \citetalias{kumar24} RVs are shown in yellow, with an additional \textsc{TheJoker} model fitting all data, including \citetalias{kumar24}, is shown in grey.}
         \label{fig:rvs}
   \end{figure}

The main result is the confirmation that this pAGB star indeed belongs to a binary system (see Fig. \ref{fig:rvs}), a first for a pAGB in a GC. Its orbital period is well constrained with a value of $34.04\pm0.05$ days. This value is consistent with one of the possible periods proposed by \citetalias{kumar24} of 39.12 $\pm$ 4.65 d. The barycentric velocity for the system is 5.13 $\pm$ 0.69 km s$^{-1}$, which can be considered in slight tension with the literature values for the cluster of 8.9 $\pm$ 2.8 km s$^{-1}$ from \citet{salinas15} based on low-resolution spectra of two stars, and 12.6 $\pm$ 0.4 km s$^{-1}$ from \citet{monaco18} obtained via high-resolution spectroscopy of 4 stars. The inconsistency can be likely attributed to the different instruments used in the different studies, but also to the fact that the cluster is undergoing tidal disruption \citep[e.g.][]{carballo20}, and to the very high binary fraction of the cluster \citep[e.g.][]{milone12} which is expected to bias the RV measurements. The semi-amplitude of the RV has a value of $K_1$=24.85 $\pm$ 1.24 km s$^{-1}$, while for the orbit eccentricity we find 0.119 $\pm$ 0.04. The posterior distribution for these quantities can be seen in Fig. \ref{fig:corner}. In addition to \textsc{TheJoker} analysis, we also performed a Markov chain Monte Carlo fit to the RV data as described in \citet{marcadon26}. Both approaches give consistent results.

With these orbital parameters we can calculate the mass function of the system,
\begin{equation}
f(m)= \frac{P_{\rm orb} K_1}{2\pi G}(1-e^2)^{\frac{3}{2}}= \frac{M_2^3\sin^3i}{(M_1+M_2)^2}=0.053\pm0.008 M_{\odot}\sin^3 i ,
\end{equation}
which is also related to the masses of both components of the binary, as well as the orbital inclination, $i$. 

Having the mass function, we can estimate the mass of the companion if values can be given to both the primary mass and $i$. Given the particular evolutionary phase these stars are going through, masses of pAGB stars are well constrained, both theoretically \citep{miller16} and observationally, since their masses are expected to be very similar to those of white dwarfs \citep{kilic18}. \citetalias{kumar24} based on a comparison with evolutionary tracks \citep{miller16, moehler19} derived a mass between 0.51 and 0.55 M$_{\odot}$ which we also adopt in this work as 0.53 $\pm$ 0.02 M$_{\odot}$. A minimum mass for the companion is obtained when assuming an edge-on system ($i=90\deg$) resulting in a secondary mass $M_2>0.343$ M$_{\odot}$. Alternatively, in the presence of a circumbinary disk \citep[e.g.][but see Sect. \ref{sec:concl}]{hillen17}, obscuration would preclude visibility of edge-on systems, and we can also assume $i=75\deg$ following \citet{oomen18}. In this case, the minimum mass for the secondary would only slightly increase to $M_2>0.36$ M$_{\odot}$.

Additionally, it is possible to calculate the projected semi-major axis of the orbit, 
\begin{equation}
    a_1 \sin(i) = \frac{P_{orb}K_1}{2\pi} \sqrt{1-e^2} =0.077\pm0.004\,\mathrm{AU},
\end{equation}
which is well within the typical radius of AGB stars \citep[e.g.][]{wittkowski17}, pointing to a past common envelope evolution.

The two velocities from \citetalias{kumar24} were excluded from the previous analysis since they appear somewhat offset to higher values compared to the GMOS velocities. If they were to be included in the analysis, the barycentric velocity would increase to $7.18\pm0.83$ km s$^{-1}$, closer to the literature values for the cluster, albeit with a higher uncertainty. This increase in the uncertainty is also present in the orbital period with P$_{\rm orb}$=34.97 $\pm$ 0.85 days. This model is also shown in Fig. \ref{fig:rvs} in grey.

A summary of the measured properties of the target are given in Table \ref{tab:target}.
 
\section{Stellar parameters and chemical abundances}\label{sec:abundances}

We analysed the continuum-normalised GHOST spectrum of the star using the non-local thermodynamic equilibrium (NLTE) BSTAR2006 model-atmosphere grid \citep{LanzHubeny2007}, \textsc{synspec} spectrum synthesis \citep{HubenyLanz2017}, and
\textsc{rotin3} instrumental and rotational convolution.

We used the BSTAR2006 models computed with a microturbulent velocity of $\xi=2$~km s$^{-1}$, as fixed by the adopted atmosphere grid. Atomic data were taken from a customized ine list from the Vienna Atomic Line Database (VALD) appropriate for the spectral range and ionisation stages relevant to this hot star. The species identifiers were converted from the VALD/MOOG format to the \textsc{synspec} convention.

\subsection{Stellar parameters and evolutionary status}
\begin{figure}[t!]
   %\centering
   \includegraphics[width=0.43\textwidth]{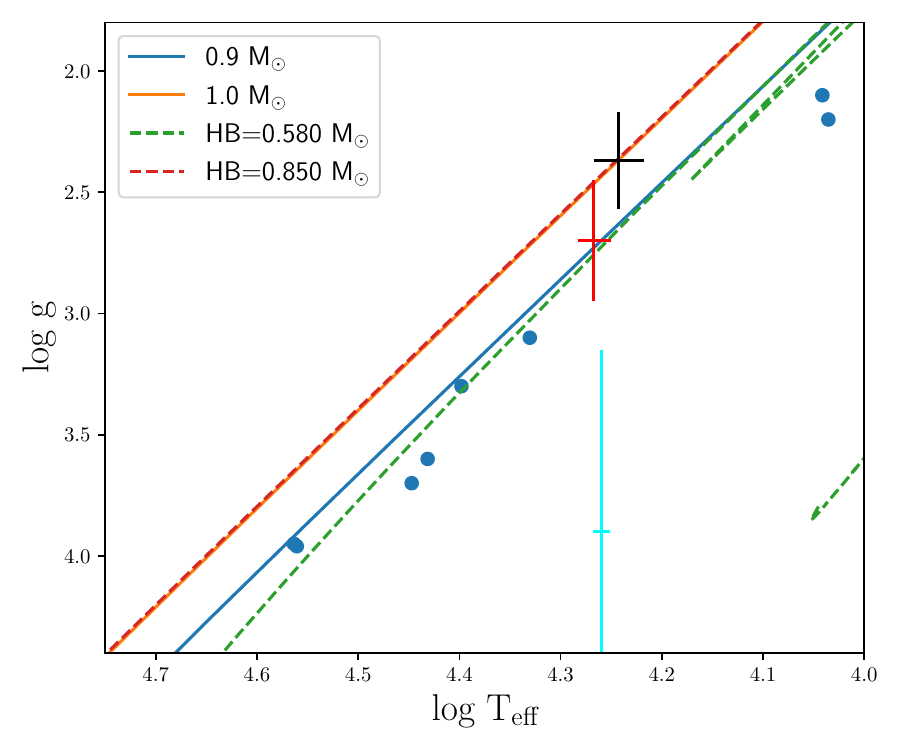}
      \caption{pAGB evolutionary tracks from \citet{miller16}, evolved from main sequence stars of 0.9 and 1.0 M$_{\odot}$ (in solid blue and orange lines, respectively); and from \cite{moehler19}, from a 0.85 M$_{\odot}$ main sequence star, but with ZAHB masses of 0.85 and 0.58 M$_{\odot}$ (in dashed lines). Our measurement of $\log g$ and $T_{\rm eff}$ is shown by a red symbol, while that of \citetalias{kumar24} is shown in black, and \citet{feeser26}, in cyan. The blue dots are 8 pAGB in GCs from the sample of \citet{dixon24}.}
         \label{fig:evolution}
   \end{figure}

Stellar parameters were determined by analysing the profiles of Balmer lines, and interpreting the behaviour of Si lines. The full explanation of the analysis is given in Appendix \ref{appendix:stellar}. The measured  effective temperature is $T_{\rm eff}$=18\,500 $\pm$ 700 K, while the surface gravity is $\log g$=2.70 $\pm$ 0.25. These values differ from the \citetalias{kumar24}
spectroscopic solution, $T_{\rm eff}=17\,000^{+400}_{-700}$ K and
$\log g=2.37\pm0.20$. The direct comparison added to Fig. C.1 shows that the \citetalias{kumar24} parameter pair produces systematically narrower gravity-sensitive H$\beta$ wings than observed, whereas at $T_{\rm eff}=17\,000$ K the Balmer locus favours $\log g\simeq2.7$. The higher gravity is independently supported by the \citetalias{kumar24} radius: combining $R=4.61\pm0.14\,R_\odot$ with the expected pAGB mass of $0.51$--$0.55\,M_\odot$ gives $\log g=2.82$--2.85. Recently, \citet{feeser26} obtained $T_{\rm eff}=18\,171^{+351}_{-353}$ K and
$\log g=3.9^{+0.7}_{-0.8}$ from SED fitting, with the latter value being much larger than those predicted by the models and those measured in GC pAGBs (see Fig. \ref{fig:evolution}).

These stellar parameters allow us to revisit the evolutionary status of the star. Fig. \ref{fig:evolution} shows a comparison with the pAGB evolutionary tracks of \citet{miller16} and \citet{moehler19}. The revised $T_{\rm eff}$, luminosity, and $\log g$ provide a better match to the pAGB evolutionary tracks. In particular, the star lies close to the \citet{miller16} track evolved from a 0.9 M$_{\odot}$ main-sequence star with Z=0.001, and corresponding to a pAGB mass of 0.53 M$_{\odot}$. The stellar parameters are also in very good agreement to the \citet{moehler19} track with [Fe/H]=--1.0 and evolved from a 0.58 M$_{\odot}$ zero age horizontal branch (ZAHB) star ( green dashed line in Fig.\ref{fig:evolution}). The higher $\log g$ compared to the one measured by \citetalias{kumar24} puts the pAGB much closer to the typical $\log g$ of pAGB in other GCs \citep[in blue markers in  Fig. \ref{fig:evolution} from][]{dixon24}, despite this, the estimated mass for the pAGB remains unchanged from the value estimated by \citetalias{kumar24}.

\subsection{Chemical abundances}

\begin{table}[t!]
\centering
\caption{Element abundances determined for the GHOST post-AGB star.  Abundances are expressed as $\log\epsilon({\rm X})=12+\log(n_{\rm X}/n_{\rm H})$ and $[\mathrm{X/H}]=\log(n_{\rm X}/n_{\rm H})_\star-
\log(n_{\rm X}/n_{\rm H})_\odot$.  Solar abundances are from
\citet{AsplundEtAl2009}.  The quantity $\sigma_{\rm L}$ is the adopted internal line-to-line/profile dispersion and $N$ is the number of accepted lines entering each abundance.}%  Atmospheric-parameter and model systematics are discussed separately and are not included in $\sigma_{\rm L}$.}
\label{tab:abundances}
\begin{tabular}{lrrrrrr}
\hline
\hline
Element & $\log\epsilon_\odot$ & $\log\epsilon_\star$ &
$[\mathrm{X/H}]$ & $\sigma_{\rm L}$ & $[\mathrm{X/Fe}]$ & $N$\\
\hline
C  & 8.43 & 7.52 & $-0.92$ & 0.19 & $-0.07$ & 8\\
N  & 7.83 & 7.59 & $-0.24$ & 0.08 & $+0.61$ & 12\\
O  & 8.69 & 8.37 & $-0.32$ & 0.18 & $+0.53$ & 16\\
Al & 6.45 & 5.80 & $-0.66$ & 0.26 & $+0.19$ & 4\\
Si & 7.51 & 7.31 & $-0.20$ & 0.12 & $+0.65$ & 2\\
Fe & 7.50 & 6.65 & $-0.85$ & 0.17 & $0.00$  & 4\\
\hline
\end{tabular}
\end{table}

The stellar parameters measured in the previous section were used to measure chemical abundances via direct profile fitting.  A complete explanation of the procedure is given in Appendix \ref{appendix:abundances}. Chemical abundances were measured for C, N, O, Al, Si and Fe (see Table \ref{tab:abundances}). No transitions associated to s-process elements were detected (see discussion in Appendix \ref{appendix:abundances}). 

The measured abundances broadly agree with those measured by \citetalias{kumar24} (see Fig. \ref{fig:abundances}), with the most notable differences being the abundances of C ($-0.19$ dex) and Al ($+0.18$ dex), where the differences are between our measurements and theirs. Additionally, the iron abundance is 0.15 dex lower than \citetalias{kumar24}, but in excellent agreement with \citet{monaco18}, [Fe/H]=--0.89 $\pm$ 0.08, who measured it from 4 red giant branch (RGB) stars in the cluster.

\citetalias{kumar24} used the C/CNO and N/O diagnostics to argue that this star had undergone a third dredge-up.  Our reanalysis gives
$\log\{\mathrm{C/(C+N+O)}\}=-0.97$ and $\log(\mathrm{N/O})=-0.78$.  The N/O ratio remains similar to that obtained
by \citetalias{kumar24}., whereas the lower carbon abundance found here implies a smaller carbon contribution to the total CNO abundance.  Consequently, the revised CNO pattern remains compatible with an evolved, CNO-processed atmosphere, but provides less support for substantial carbon enrichment by third dredge-up. 

\section{Binary population synthesis}\label{sec:synthesis}
\begin{figure}[t!]
   %\centering
   \includegraphics[width=0.5\textwidth]{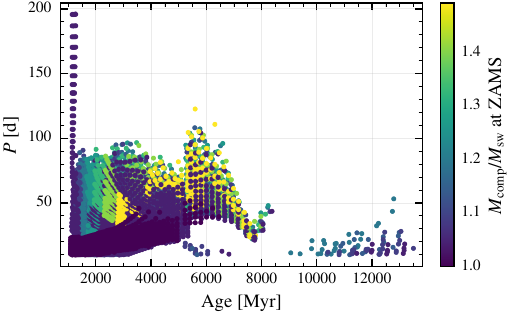}
      \caption{Distribution of age (time since formation) at which the initially less massive star in a binary system evolves towards the pAGB stage, against the orbital period after the transition. The colour map shows the ZAMS companion mass ($M_{\rm comp}$) to pAGB mass (or stage switching star, hence $M_{\rm SW}$) ratio. This figure corresponds to the $\alpha_{\rm CE}=3.0$ configuration, as described in Appendix \ref{appendix:compas}, and includes a subset of systems that matches cluster E3's age estimates.}
         \label{fig:alpha3}
   \end{figure}
   
Perhaps the most remarkable characteristic of the system is its short orbital period. An inspection of the 33 binary pAGB studied by \citet{oomen18} shows that no binary pAGB has an orbital period shorter than 100 days (Fig. \ref{fig:periods}).

In order to constrain the origin of such a short orbital period we conducted a number of simulations using the rapid binary population synthesis code COmpact Object Mergers: Population Astrophysics \& Statistics \citep[\textsc{compas},][]{riley22,compas25}. Details of the \textsc{compas} simulations are given in Appendix \ref{appendix:compas}, including what we define as pAGB in \textsc{compas}, and the grid of input parameters. Among the latter, perhaps the most important component is $\alpha_{\rm CE}$, the efficiency with which orbital energy is converted into mechanical energy to expel the common envelope \citep{webbink84} formed during episodes of dynamically unstable mass transfer. \textsc{compas} models were run with  $\alpha_{CE}$=0.1, 1.0, 3.0 and 5.0.

For $\alpha_{\rm CE}$=0.1, a value motivated by the results of \citet{Zorotovic2010}, there are no systems older than 1.1 Gyr with periods in the relevant range (between 10 and 100 days). Furthermore, all young systems have orbital periods longer ($\geq60$ d) than the observational estimates. Another problem is that these systems have $q\sim1$ at the ZAMS, with the primary star being $\sim1.8\,M_\odot$. Low values of  $\alpha_{\rm CE}$ are therefore incompatible with the orbital period, and with the age limits imposed by the cluster membership.

Models with larger values for $\alpha_{\rm CE}$ show an increasing number of systems with older ages and orbital periods compatible with the measured period. When the initially more massive star in the system becomes the pAGB, the number of systems increases from $\sim6\,000$ at $\alpha_{\rm CE}=1$ to $\approx 18\,000$ at $\alpha_{\rm CE}=5$, with system ages ranging from 1 to 6 Gyr at $\alpha_{\rm CE}=1$, and from 1 to 7 Gyr at $\alpha_{\rm CE}=5$. Despite better agreement with the measured orbital period, all these systems are far from the cluster's age estimate. Still, they would remain viable scenarios if the pAGB were not a member of E3.

 %Larger values of $\alpha_{\rm CE}$ show an increasing number of compatible systems and maximum age for increasing $\alpha_{\rm CE}$ values in cases where the primary becomes a pAGB with orbital period in the relevant range. We go from about $6\,000$ systems at $\alpha_{\rm CE}=1$, to $\approx 18\,000$ at $\alpha_{\rm CE}=5$. The total age range covered by these systems goes from 1-6 Gyr to 1-7 Gyr. Additionally, most companions are main sequence stars, though those with $M>0.7\,M_\odot$ are often found in systems with $P_{\rm orb} >70$ d. Still, all these systems are too young if we compare them against E3's age estimates (10--13 Gyr), but remain possible if the pAGB were not to be a member of the cluster.

While the previous scenario requires the initially more massive star to become the present-day pAGB, a different situation arises when the initially less massive companion instead occupies this role. In this case, the number of systems with periods similar to the observed one also increases when $\alpha_{\rm CE}$ increases and, more importantly, systems with ages greater than 10 Gyr are found (see Fig. \ref{fig:alpha3}). In this scenario, the companions to the pAGB are mostly helium white dwarfs (HeWDs, see Fig. \ref{fig:types}), hinting at a mass transfer episode at an earlier evolutionary stage, which raises questions about our chosen physical assumptions (e.g., the value of $\alpha_{\rm CE}$ or mass loss through a possible circumbinary disc).  The detailed evolution mode in \textsc{compas} shows that at least some pAGB+HeWD systems form via an early stable Roche-lobe overflow (HeWD formation) plus a later common envelope when the HeWD has already cooled for a couple of Gyr (pAGB formation), making it undetectable by SED fitting. Despite the limitations of the modelling, we propose this as the most likely formation scenario for this pAGB.

% When \textbf{the initially less massive star} becomes the pAGB in the system, the numbers also increase with $\alpha_{\rm CE}$ but, unlike the previous scenario, starting at $\alpha_{\rm CE}=3$ there are systems with age greater than 10 Gyr and periods in the range 10-50 d (see Fig. \ref{fig:alpha3}). However, in this scenario the companions are typically Helium white dwarfs (HeWDs, see Fig. \ref{fig:types}), hinting at mass transfer episodes at earlier evolutionary stages. This complicates the evolutionary history and raises questions about the chosen physical assumptions (e.g., $\alpha_{\rm CE}$ or mass loss through a possible circumbinary disc). Note, however, that the detailed evolution mode in \textsc{compas} shows that at least some pAGB+HeWD systems are formed by an early stable Roche-lobe overflow (HeWD formation) plus a later common envelope (pAGB formation), when the HeWD has already cooled for a couple of Gyr, making it undetectable by SED fitting. Despite the limitations of the modelling, we propose this as the most likely formation scenario of this pAGB.
  
\section{Discussion and conclusions}\label{sec:concl}

One of the defining characteristics of pAGB in binary systems is the existence of circumbinary discs \citep[e.g.][]{deruyter06}, but the pAGB studied in this paper shows little to no evidence of such occurrence. The SED fitting presented by \citetalias{kumar24} shows no NIR or MIR excess, and the star colours, with 2MASS $H-K$=0.01 \citep{skrutskie06}, and WISE $W1-W3$=0.59 \citep{wright19}, do not satisfy the criteria of \citet{kluska22} for the existence of a full disk, although a ``transition disk'' \citep[those with large inner cavities, e.g.][]{andrews11} may still be possible.

The existence of circumbinary disks also changes the surface chemistry of the pAGB star. While refractory elements will be trapped in the dusty disc, volatile elements will reaccrete on to the surface of the pAGB, making for a relative depletion of the measured abundances of refractory elements \citep[e.g.][]{oomen19}. This depletion is quantified by volatile-to-refractory ratios such as [Zn/Ti] or [S/Ti] \citep{oomen18}. Unfortunately, at this high effective temperature there are no accessible Zn or Ti transitions in the optical spectrum, while the four usable
S\,\textsc{ii} lines yield multiplet-dependent abundances differing by
$\sim$0.9 dex. We therefore do not adopt a sulphur abundance; \citetalias{kumar24} encountered a similar inconsistency and reported only an upper limit.
But a hint is given by the behaviour of abundance of Fe, a refractory element itself, which is essentially identical to that of RGB stars in the cluster \citep{monaco18}, even though iron depletion can be as high as a few dex in some binary pAGB \citep[e.g.][]{mohorian25}. 

The compound evidence therefore points against the existence of any circumbinary disc around this system. Since many pAGB binaries have been discovered thanks to the fingerprints of circumbinary discs on the pAGB colours \citep[e.g.][]{szczerba07}, the pAGB in E3 may indicate an hitherto hidden population of binary pAGB, where no disc was formed, perhaps by the complex formation of the system (Sect. \ref{sec:synthesis}) and the proximity of its companion, or where it was too short-lived to change the photospheric abundances.

The BPS models indicate that multiple mass transfer episodes are required in order to explain the short orbital period.  A complete understanding of the system, including the apparent absence of any circumbinary disc, will require more detail modelling beyond what is possible with BPS.

\begin{acknowledgements} 
 R.S. acknowledges support from the European Research Council (ERC) under the European Union’s Horizon 2020 research and innovation program (grant agreement No. 951549 - UniverScale). C.S. acknowledges support from ANID FONDECYT POSTDOCTORADO N°3240157. N.R.S. acknowledges support by the Korea Astronomy and Space Science Institute under the R\&D program (Project No. 2026-1-831-02) supervised by the Korea AeroSpace Administration. Support for M.C. is provided by ANID’s FONDECYT Regular grant \#1231637 and ANID’s Basal project FB210003. A.E. received the support of a fellowship from "La Caixa" Foundation (ID 100010434) with fellowship code LCF/BQ/PI23/11970031. Claude Code AI was used to adapt the CERES pipeline to reduce BESO spectra. Based on observations obtained at the international Gemini Observatory, a program of NOIRLab, which is managed by the Association of Universities for Research in Astronomy (AURA) under a cooperative agreement with the National Science Foundation. on behalf of the Gemini Observatory partnership: the National Science Foundation (United States), National Research Council (Canada), Agencia Nacional de Investigaci\'{o}n y Desarrollo (Chile), Ministerio de Ciencia, Tecnolog\'{i}a e Innovaci\'{o}n (Argentina), Minist\'{e}rio da Ci\^{e}ncia, Tecnologia, Inova\c{c}\~{o}es e Comunica\c{c}\~{o}es (Brazil), and Korea Astronomy and Space Science Institute (Republic of Korea).
\end{acknowledgements}

\bibliographystyle{aa}
\bibliography{pagb}

@ARTICLE{hook04,
       author = {{Hook}, I.~M. and {J{\o}rgensen}, Inger and {Allington-Smith}, J.~R. and {Davies}, R.~L. and {Metcalfe}, N. and {Murowinski}, R.~G. and {Crampton}, D.},
        title = "{The Gemini-North Multi-Object Spectrograph: Performance in Imaging, Long-Slit, and Multi-Object Spectroscopic Modes}",
      journal = {\pasp},
         year = 2004,
        month = may,
       volume = {116},
       number = {819},
        pages = {425-440},
          doi = {10.1086/383624},
       adsurl = {https://ui.adsabs.harvard.edu/abs/2004PASP..116..425H}
}

@ARTICLE{dragons,
       author = {{Labrie}, K. and {Simpson}, C. and {Cardenes}, R. and {Turner}, J. and {Soraisam}, M. and {Quint}, B. and {Oberdorf}, O. and {Placco}, V.~M. and {Berke}, D. and {Smirnova}, O. and {Conseil}, S. and {Vacca}, W.~D. and {Thomas-Osip}, J.},
        title = "{DRAGONS-A Quick Overview}",
      journal = {Research Notes of the American Astronomical Society},
         year = 2023,
        month = oct,
       volume = {7},
       number = {10},
          eid = {214},
        pages = {214},
          doi = {10.3847/2515-5172/ad0044},
archivePrefix = {arXiv},
       eprint = {2310.03048},
 primaryClass = {astro-ph.IM},
       adsurl = {https://ui.adsabs.harvard.edu/abs/2023RNAAS...7..214L}
}

@ARTICLE{thejoker,
       author = {{Price-Whelan}, Adrian M. and {Hogg}, David W. and {Foreman-Mackey}, Daniel and {Rix}, Hans-Walter},
        title = "{The Joker: A Custom Monte Carlo Sampler for Binary-star and Exoplanet Radial Velocity Data}",
      journal = {\apj},
         year = 2017,
        month = mar,
       volume = {837},
       number = {1},
          eid = {20},
        pages = {20},
          doi = {10.3847/1538-4357/aa5e50},
archivePrefix = {arXiv},
       eprint = {1610.07602},
 primaryClass = {astro-ph.SR},
       adsurl = {https://ui.adsabs.harvard.edu/abs/2017ApJ...837...20P}
}

@ARTICLE{kalari24,
       author = {{Kalari}, V.~M. and {Diaz}, R.~J. and {Robertson}, G. and {McConnachie}, A. and {Ireland}, M. and {Salinas}, R. and {Young}, P. and {Simpson}, C. and {Hayes}, C. and {Nielsen}, J. and {Burley}, G. and {Pazder}, J. and {Gomez-Jimenez}, M. and {Martioli}, E. and {Howell}, S.~B. and {Jeong}, M. and {Juneau}, S. and {Ruiz-Carmona}, R. and {Margheim}, S. and {Sheinis}, A. and {Anthony}, A. and {Baker}, G. and {Berg}, T.~A.~M. and {Cao}, T. and {Chapin}, E. and {Chin}, T. and {Chiboucas}, K. and {Churilov}, V. and {Deibert}, E. and {Densmore}, A. and {Dunn}, J. and {Edgar}, M.~L. and {Heo}, J. and {Henderson}, D. and {Farrell}, T. and {Font}, J. and {Firpo}, V. and {Fuentes}, J. and {Labrie}, K. and {Lambert}, S. and {Lawrence}, J. and {Lothrop}, J. and {McDermid}, R. and {Miller}, B.~W. and {Perez}, G. and {Placco}, V.~M. and {Prado}, P. and {Quiroz}, C. and {Ramos}, F. and {Rutten}, R. and {Silva}, K.~M.~G. and {Thomas-Osip}, J. and {Urrutia}, C. and {Vacca}, W.~D. and {Venn}, K. and {Waller}, F. and {Waller}, L. and {White}, M. and {Xu}, S. and {Zhelem}, R.},
        title = "{Gemini High-resolution Optical Spectrograph (GHOST) at Gemini South: Instrument Performance and Integration, First Science, and Next Steps}",
      journal = {\aj},
         year = 2024,
        month = nov,
       volume = {168},
       number = {5},
          eid = {208},
        pages = {208},
          doi = {10.3847/1538-3881/ad72ed},
archivePrefix = {arXiv},
       eprint = {2409.05855},
 primaryClass = {astro-ph.IM},
       adsurl = {https://ui.adsabs.harvard.edu/abs/2024AJ....168..208K}
}

@ARTICLE{placco24,
       author = {{Placco}, Vinicius M. and {Herrera}, David and {Merino}, Brian M. and {US National Gemini Office} and {Hirst}, Paul and {Labrie}, Kathleen and {Simpson}, Chris and {Turner}, James and {Vacca}, William D. and {Gemini Science User Support Department} and {Deibert}, Emily and {Diaz}, Ruben and {Heo}, Jeong-Eun and {Kalari}, Venu and {Reggiani}, Henrique and {Rodriguez}, Cinthya and {Ruiz-Carmona}, Roque and {Thomas-Osip}, Joanna and {GHOST Instrument Team}},
        title = "{GHOST Reduced Data Products for the Gemini Observatory Community and Beyond}",
      journal = {Research Notes of the American Astronomical Society},
         year = 2024,
        month = dec,
       volume = {8},
       number = {12},
          eid = {312},
        pages = {312},
          doi = {10.3847/2515-5172/ad9f5f},
archivePrefix = {arXiv},
       eprint = {2412.13239},
 primaryClass = {astro-ph.IM},
       adsurl = {https://ui.adsabs.harvard.edu/abs/2024RNAAS...8..312P}
}

@ARTICLE{carballo20,
       author = {{Carballo-Bello}, Julio A. and {Salinas}, Ricardo and {Piatti}, Andr{\'e}s E.},
        title = "{The disruption of the low-mass globular cluster E 3}",
      journal = {\mnras},
         year = 2020,
        month = dec,
       volume = {499},
       number = {2},
        pages = {2157-2161},
          doi = {10.1093/mnras/staa2864},
archivePrefix = {arXiv},
       eprint = {2009.11320},
 primaryClass = {astro-ph.GA},
       adsurl = {https://ui.adsabs.harvard.edu/abs/2020MNRAS.499.2157C}
}

@ARTICLE{salinas15,
       author = {{Salinas}, Ricardo and {Strader}, Jay},
        title = "{No Evidence for Multiple Stellar Populations in the Low-mass Galactic Globular Cluster E 3}",
      journal = {\apj},
         year = 2015,
        month = aug,
       volume = {809},
       number = {2},
          eid = {169},
        pages = {169},
          doi = {10.1088/0004-637X/809/2/169},
archivePrefix = {arXiv},
       eprint = {1506.00637},
 primaryClass = {astro-ph.GA},
       adsurl = {https://ui.adsabs.harvard.edu/abs/2015ApJ...809..169S}
}

@ARTICLE{monaco18,
       author = {{Monaco}, L. and {Villanova}, S. and {Carraro}, G. and {Mucciarelli}, A. and {Moni Bidin}, C.},
        title = "{Chemical abundance analysis of red giant branch stars in the globular cluster E3}",
      journal = {\aap},
         year = 2018,
        month = sep,
       volume = {616},
          eid = {A181},
        pages = {A181},
          doi = {10.1051/0004-6361/201832979},
archivePrefix = {arXiv},
       eprint = {1806.03327},
 primaryClass = {astro-ph.SR},
       adsurl = {https://ui.adsabs.harvard.edu/abs/2018A&A...616A.181M}
}

@ARTICLE{kumar24,
       author = {{Kumar}, R. and {Moharana}, A. and {Piridi}, S. and {Pradhan}, A.~C. and {He{\l}miniak}, K.~G. and {Ikonnikova}, N. and {Dodin}, A. and {Szczerba}, R. and {Giersz}, M. and {Ojha}, D.~K. and {Samal}, M.~R.},
        title = "{Discovery of a hot post-AGB star in Galactic globular cluster E3}",
      journal = {\aap},
         year = 2024,
        month = may,
       volume = {685},
          eid = {L6},
        pages = {L6},
          doi = {10.1051/0004-6361/202449777},
archivePrefix = {arXiv},
       eprint = {2403.12907},
 primaryClass = {astro-ph.SR},
       adsurl = {https://ui.adsabs.harvard.edu/abs/2024A&A...685L...6K},
       shorthand = {K24} 
}

@ARTICLE{moehler19,
       author = {{Moehler}, S. and {Landsman}, W.~B. and {Lanz}, T. and {Miller Bertolami}, M.~M.},
        title = "{Hot UV-bright stars of galactic globular clusters}",
      journal = {\aap},
         year = 2019,
        month = jul,
       volume = {627},
          eid = {A34},
        pages = {A34},
          doi = {10.1051/0004-6361/201935694},
archivePrefix = {arXiv},
       eprint = {1905.06718},
 primaryClass = {astro-ph.SR},
       adsurl = {https://ui.adsabs.harvard.edu/abs/2019A&A...627A..34M}
}

@ARTICLE{miller16,
       author = {{Miller Bertolami}, Marcelo Miguel},
        title = "{New models for the evolution of post-asymptotic giant branch stars and central stars of planetary nebulae}",
      journal = {\aap},
         year = 2016,
        month = apr,
       volume = {588},
          eid = {A25},
        pages = {A25},
          doi = {10.1051/0004-6361/201526577},
archivePrefix = {arXiv},
       eprint = {1410.1679},
 primaryClass = {astro-ph.SR},
       adsurl = {https://ui.adsabs.harvard.edu/abs/2016A&A...588A..25M}
}

@ARTICLE{oomen18,
       author = {{Oomen}, Glenn-Michael and {Van Winckel}, Hans and {Pols}, Onno and {Nelemans}, Gijs and {Escorza}, Ana and {Manick}, Rajeev and {Kamath}, Devika and {Waelkens}, Christoffel},
        title = "{Orbital properties of binary post-AGB stars}",
      journal = {\aap},
         year = 2018,
        month = dec,
       volume = {620},
          eid = {A85},
        pages = {A85},
          doi = {10.1051/0004-6361/201833816},
archivePrefix = {arXiv},
       eprint = {1810.01842},
 primaryClass = {astro-ph.SR},
       adsurl = {https://ui.adsabs.harvard.edu/abs/2018A&A...620A..85O}
}

@ARTICLE{kilic18,
       author = {{Kilic}, Mukremin and {Hambly}, N.~C. and {Bergeron}, P. and {Genest-Beaulieu}, C. and {Rowell}, N.},
        title = "{Gaia reveals evidence for merged white dwarfs}",
      journal = {\mnras},
         year = 2018,
        month = sep,
       volume = {479},
       number = {1},
        pages = {L113-L117},
          doi = {10.1093/mnrasl/sly110},
archivePrefix = {arXiv},
       eprint = {1805.01227},
 primaryClass = {astro-ph.SR},
       adsurl = {https://ui.adsabs.harvard.edu/abs/2018MNRAS.479L.113K}
}

@ARTICLE{hillen17,
       author = {{Hillen}, M. and {Van Winckel}, H. and {Menu}, J. and {Manick}, R. and {Debosscher}, J. and {Min}, M. and {de Wit}, W.-J. and {Verhoelst}, T. and {Kamath}, D. and {Waters}, L.~B.~F.~M.},
        title = "{A mid-IR interferometric survey with MIDI/VLTI: resolving the second-generation protoplanetary disks around post-AGB binaries}",
      journal = {\aap},
         year = 2017,
        month = mar,
       volume = {599},
          eid = {A41},
        pages = {A41},
          doi = {10.1051/0004-6361/201629161},
archivePrefix = {arXiv},
       eprint = {1610.09930},
 primaryClass = {astro-ph.SR},
       adsurl = {https://ui.adsabs.harvard.edu/abs/2017A&A...599A..41H}
}

@INPROCEEDINGS{steiner06,
       author = {{Steiner}, Ingo and {Seifert}, Walter and {Stahl}, Otmar and {Lemke}, Roland and {Chini}, Rolf and {Appenzeller}, Immo},
        title = "{BESO: a high-resolution spectrograph for the Hexapod-Telescope}",
    booktitle = {Ground-based and Airborne Instrumentation for Astronomy},
         year = 2006,
       editor = {{McLean}, Ian S. and {Iye}, Masanori},
       series = {Society of Photo-Optical Instrumentation Engineers (SPIE) Conference Series},
       volume = {6269},
        month = jun,
          eid = {62692W},
        pages = {62692W},
          doi = {10.1117/12.671120},
       adsurl = {https://ui.adsabs.harvard.edu/abs/2006SPIE.6269E..2WS}
}

@ARTICLE{dixon24,
       author = {{Dixon}, William V.},
        title = "{Evidence of Third Dredge-up in Post-AGB Stars in Galactic Globular Clusters}",
      journal = {\apj},
         year = 2024,
        month = jan,
       volume = {961},
       number = {1},
          eid = {24},
        pages = {24},
          doi = {10.3847/1538-4357/ad0a89},
archivePrefix = {arXiv},
       eprint = {2311.04775},
 primaryClass = {astro-ph.SR},
       adsurl = {https://ui.adsabs.harvard.edu/abs/2024ApJ...961...24D}
}

@article{LanzHubeny2007,
  author = {Lanz, T. and Hubeny, I.},
  title = {A Grid of NLTE Line-blanketed Model Atmospheres of Early B-Type
           Stars},
  journal = {\apjs},
  volume = {169},
  pages = {83--104},
  year = {2007},
  doi = {10.1086/511270}
}

@misc{HubenyLanz2017,
  author = {Hubeny, I. and Lanz, T.},
  title = {A brief introductory guide to TLUSTY and SYNSPEC},
  year = {2017},
  eprint = {1706.01859},
  archivePrefix = {arXiv}
}

@article{AsplundEtAl2009,
  author = {Asplund, M. and Grevesse, N. and Sauval, A. J. and Scott, P.},
  title = {The chemical composition of the Sun},
  journal = {\araa},
  volume = {47},
  pages = {481--522},
  year = {2009},
  doi = {10.1146/annurev.astro.46.060407.145222}
}

@ARTICLE{milone12,
       author = {{Milone}, A.~P. and {Piotto}, G. and {Bedin}, L.~R. and {Aparicio}, A. and {Anderson}, J. and {Sarajedini}, A. and {Marino}, A.~F. and {Moretti}, A. and {Davies}, M.~B. and {Chaboyer}, B. and {Dotter}, A. and {Hempel}, M. and {Mar{\'\i}n-Franch}, A. and {Majewski}, S. and {Paust}, N.~E.~Q. and {Reid}, I.~N. and {Rosenberg}, A. and {Siegel}, M.},
        title = "{The ACS survey of Galactic globular clusters. XII. Photometric binaries along the main sequence}",
      journal = {\aap},
         year = 2012,
        month = apr,
       volume = {540},
          eid = {A16},
        pages = {A16},
          doi = {10.1051/0004-6361/201016384},
archivePrefix = {arXiv},
       eprint = {1111.0552},
 primaryClass = {astro-ph.SR},
       adsurl = {https://ui.adsabs.harvard.edu/abs/2012A&A...540A..16M}
}

@ARTICLE{wittkowski17,
       author = {{Wittkowski}, M. and {Hofmann}, K.-H. and {H{\"o}fner}, S. and {Le Bouquin}, J.~B. and {Nowotny}, W. and {Paladini}, C. and {Young}, J. and {Berger}, J.-P. and {Brunner}, M. and {de Gregorio-Monsalvo}, I. and {Eriksson}, K. and {Hron}, J. and {Humphreys}, E.~M.~L. and {Lindqvist}, M. and {Maercker}, M. and {Mohamed}, S. and {Olofsson}, H. and {Ramstedt}, S. and {Weigelt}, G.},
        title = "{Aperture synthesis imaging of the carbon AGB star R Sculptoris. Detection of a complex structure and a dominating spot on the stellar disk}",
      journal = {\aap},
         year = 2017,
        month = may,
       volume = {601},
          eid = {A3},
        pages = {A3},
          doi = {10.1051/0004-6361/201630214},
archivePrefix = {arXiv},
       eprint = {1702.02574},
 primaryClass = {astro-ph.SR},
       adsurl = {https://ui.adsabs.harvard.edu/abs/2017A&A...601A...3W}
}

@ARTICLE{vanwinckel03,
       author = {{van Winckel}, Hans},
        title = "{Post-AGB Stars}",
      journal = {\araa},
         year = 2003,
        month = jan,
       volume = {41},
        pages = {391-427},
          doi = {10.1146/annurev.astro.41.071601.170018},
       adsurl = {https://ui.adsabs.harvard.edu/abs/2003ARA&A..41..391V}
}

@ARTICLE{vanwinckel18,
       author = {{Van Winckel}, Hans},
        title = "{Binary post-AGB stars as tracers of stellar evolution}",
      journal = {arXiv e-prints},
         year = 2018,
        month = sep,
          eid = {arXiv:1809.00871},
        pages = {arXiv:1809.00871},
          doi = {10.48550/arXiv.1809.00871},
archivePrefix = {arXiv},
       eprint = {1809.00871},
 primaryClass = {astro-ph.SR},
       adsurl = {https://ui.adsabs.harvard.edu/abs/2018arXiv180900871V}
}

@ARTICLE{mohorian25,
       author = {{Mohorian}, Maksym and {Kamath}, Devika and {Menon}, Meghna and {Van Winckel}, Hans and {Jian}, Mingjie and {Amarsi}, Anish M. and {Andrych}, Kateryna},
        title = "{Abundance analysis of chemically depleted post-AGB/post-RGB binaries with faint discs}",
      journal = {\pasa},
         year = 2025,
        month = oct,
       volume = {42},
          eid = {e151},
        pages = {e151},
          doi = {10.1017/pasa.2025.10117},
archivePrefix = {arXiv},
       eprint = {2510.16691},
 primaryClass = {astro-ph.SR},
       adsurl = {https://ui.adsabs.harvard.edu/abs/2025PASA...42..151M}
}

@ARTICLE{kluska22,
       author = {{Kluska}, J. and {Van Winckel}, H. and {Copp{\'e}e}, Q. and {Oomen}, G.-M. and {Dsilva}, K. and {Kamath}, D. and {Bujarrabal}, V. and {Min}, M.},
        title = "{A population of transition disks around evolved stars: Fingerprints of planets. Catalog of disks surrounding Galactic post-AGB binaries}",
      journal = {\aap},
         year = 2022,
        month = feb,
       volume = {658},
          eid = {A36},
        pages = {A36},
          doi = {10.1051/0004-6361/202141690},
archivePrefix = {arXiv},
       eprint = {2201.13155},
 primaryClass = {astro-ph.EP},
       adsurl = {https://ui.adsabs.harvard.edu/abs/2022A&A...658A..36K}
}

@ARTICLE{deruyter06,
       author = {{de Ruyter}, S. and {van Winckel}, H. and {Maas}, T. and {Lloyd Evans}, T. and {Waters}, L.~B.~F.~M. and {Dejonghe}, H.},
        title = "{Keplerian discs around post-AGB stars: a common phenomenon?}",
      journal = {\aap},
         year = 2006,
        month = mar,
       volume = {448},
       number = {2},
        pages = {641-653},
          doi = {10.1051/0004-6361:20054062},
archivePrefix = {arXiv},
       eprint = {astro-ph/0601578},
 primaryClass = {astro-ph},
       adsurl = {https://ui.adsabs.harvard.edu/abs/2006A&A...448..641D}
}

@ARTICLE{skrutskie06,
       author = {{Skrutskie}, M.~F. and {Cutri}, R.~M. and {Stiening}, R. and {Weinberg}, M.~D. and {Schneider}, S. and {Carpenter}, J.~M. and {Beichman}, C. and {Capps}, R. and {Chester}, T. and {Elias}, J. and {Huchra}, J. and {Liebert}, J. and {Lonsdale}, C. and {Monet}, D.~G. and {Price}, S. and {Seitzer}, P. and {Jarrett}, T. and {Kirkpatrick}, J.~D. and {Gizis}, J.~E. and {Howard}, E. and {Evans}, T. and {Fowler}, J. and {Fullmer}, L. and {Hurt}, R. and {Light}, R. and {Kopan}, E.~L. and {Marsh}, K.~A. and {McCallon}, H.~L. and {Tam}, R. and {Van Dyk}, S. and {Wheelock}, S.},
        title = "{The Two Micron All Sky Survey (2MASS)}",
      journal = {\aj},
         year = 2006,
        month = feb,
       volume = {131},
       number = {2},
        pages = {1163-1183},
          doi = {10.1086/498708},
       adsurl = {https://ui.adsabs.harvard.edu/abs/2006AJ....131.1163S}
}

@ARTICLE{andrews11,
       author = {{Andrews}, Sean M. and {Wilner}, David J. and {Espaillat}, Catherine and {Hughes}, A.~M. and {Dullemond}, C.~P. and {McClure}, M.~K. and {Qi}, Chunhua and {Brown}, J.~M.},
        title = "{Resolved Images of Large Cavities in Protoplanetary Transition Disks}",
      journal = {\apj},
         year = 2011,
        month = may,
       volume = {732},
       number = {1},
          eid = {42},
        pages = {42},
          doi = {10.1088/0004-637X/732/1/42},
archivePrefix = {arXiv},
       eprint = {1103.0284},
 primaryClass = {astro-ph.GA},
       adsurl = {https://ui.adsabs.harvard.edu/abs/2011ApJ...732...42A}
}

@ARTICLE{riley22,
       author = {{Riley}, Jeff and {Agrawal}, Poojan and {Barrett}, Jim W. and {Boyett}, Kristan N.~K. and {Broekgaarden}, Floor S. and {Chattopadhyay}, Debatri and {Gaebel}, Sebastian M. and {Gittins}, Fabian and {Hirai}, Ryosuke and {Howitt}, George and {Justham}, Stephen and {Khandelwal}, Lokesh and {Kummer}, Floris and {Lau}, Mike Y.~M. and {Mandel}, Ilya and {de Mink}, Selma E. and {Neijssel}, Coenraad and {Riley}, Tim and {van Son}, Lieke and {Stevenson}, Simon and {Vigna-G{\'o}mez}, Alejandro and {Vinciguerra}, Serena and {Wagg}, Tom and {Willcox}, Reinhold and {Team Compas}},
        title = "{Rapid Stellar and Binary Population Synthesis with COMPAS}",
      journal = {\apjs},
         year = 2022,
        month = feb,
       volume = {258},
       number = {2},
          eid = {34},
        pages = {34},
          doi = {10.3847/1538-4365/ac416c},
archivePrefix = {arXiv},
       eprint = {2109.10352},
 primaryClass = {astro-ph.IM},
       adsurl = {https://ui.adsabs.harvard.edu/abs/2022ApJS..258...34R}
}

@ARTICLE{compas25,
       author = {{Team Compas} and {Mandel}, Ilya and {Riley}, Jeff and {Boesky}, Adam and {Br{\v{c}}ek}, Adam and {Hirai}, Ryosuke and {Kapil}, Veome and {Lau}, Mike Y.~M. and {Merritt}, Jd and {Rodr{\'\i}guez-Segovia}, Nicol{\'a}s and {Romero-Shaw}, Isobel and {Song}, Yuzhe and {Stevenson}, Simon and {Vajpeyi}, Avi and {van Son}, L.~A.~C. and {Vigna-G{\'o}mez}, Alejandro and {Willcox}, Reinhold},
        title = "{Rapid Stellar and Binary Population Synthesis with COMPAS: Methods Paper II}",
      journal = {\apjs},
         year = 2025,
        month = sep,
       volume = {280},
       number = {1},
          eid = {43},
        pages = {43},
          doi = {10.3847/1538-4365/adf8d0},
archivePrefix = {arXiv},
       eprint = {2506.02316},
 primaryClass = {astro-ph.SR},
       adsurl = {https://ui.adsabs.harvard.edu/abs/2025ApJS..280...43T}
}

@ARTICLE{oomen19,
       author = {{Oomen}, Glenn-Michael and {Van Winckel}, Hans and {Pols}, Onno and {Nelemans}, Gijs},
        title = "{Modelling depletion by re-accretion of gas from a dusty disc in post-AGB stars}",
      journal = {\aap},
         year = 2019,
        month = sep,
       volume = {629},
          eid = {A49},
        pages = {A49},
          doi = {10.1051/0004-6361/201935853},
archivePrefix = {arXiv},
       eprint = {1908.01788},
 primaryClass = {astro-ph.SR},
       adsurl = {https://ui.adsabs.harvard.edu/abs/2019A&A...629A..49O}
}

@ARTICLE{szczerba07,
       author = {{Szczerba}, R. and {Si{\'o}dmiak}, N. and {Stasi{\'n}ska}, G. and {Borkowski}, J.},
        title = "{An evolutionary catalogue of galactic post-AGB and related objects}",
      journal = {\aap},
         year = 2007,
        month = jul,
       volume = {469},
       number = {2},
        pages = {799-806},
          doi = {10.1051/0004-6361:20067035},
archivePrefix = {arXiv},
       eprint = {astro-ph/0703717},
 primaryClass = {astro-ph},
       adsurl = {https://ui.adsabs.harvard.edu/abs/2007A&A...469..799S}
}

@ARTICLE{feeser26,
       author = {{Feeser}, S. Jean and {Ciardullo}, Robin and {Siegel}, Michael H. and {Hapich}, C.~J. and {Chaturvedi}, Akshat S.},
        title = "{Population II Post-AGB Stars. I. Methodology, Initial Results, and the First Luminosity Function}",
      journal = {arXiv e-prints},
         year = 2026,
        month = oct,
          eid = {arXiv:2610.10739},
        pages = {arXiv:2610.10739},
archivePrefix = {arXiv},
       eprint = {2610.10739},
 primaryClass = {astro-ph.SR},
       adsurl = {https://ui.adsabs.harvard.edu/abs/2026arXiv261010739F}
}

@ARTICLE{FuenteMarcos2015,
       author = {{de la Fuente Marcos}, R. and {de la Fuente Marcos}, C. and {Moni Bidin}, C. and {Ortolani}, S. and {Carraro}, G.},
        title = "{Ghosts of Milky Way's past: the globular cluster ESO 37-1 (E 3)}",
      journal = {\aap},
         year = 2015,
        month = sep,
       volume = {581},
          eid = {A13},
        pages = {A13},
          doi = {10.1051/0004-6361/201526580},
archivePrefix = {arXiv},
       eprint = {1507.01347},
 primaryClass = {astro-ph.GA},
       adsurl = {https://ui.adsabs.harvard.edu/abs/2015A&A...581A..13D}
}

@ARTICLE{Zorotovic2010,
       author = {{Zorotovic}, M. and {Schreiber}, M.~R. and {G{\"a}nsicke}, B.~T. and {Nebot G{\'o}mez-Mor{\'a}n}, A.},
        title = "{Post-common-envelope binaries from SDSS. IX: Constraining the common-envelope efficiency}",
      journal = {\aap},
         year = 2010,
        month = sep,
       volume = {520},
          eid = {A86},
        pages = {A86},
          doi = {10.1051/0004-6361/200913658},
archivePrefix = {arXiv},
       eprint = {1006.1621},
 primaryClass = {astro-ph.SR},
       adsurl = {https://ui.adsabs.harvard.edu/abs/2010A&A...520A..86Z}
}

@ARTICLE{Ropke2023,
       author = {{R{\"o}pke}, Friedrich K. and {De Marco}, Orsola},
        title = "{Simulations of common-envelope evolution in binary stellar systems: physical models and numerical techniques}",
      journal = {Living Reviews in Computational Astrophysics},
         year = 2023,
        month = dec,
       volume = {9},
       number = {1},
          eid = {2},
        pages = {2},
          doi = {10.1007/s41115-023-00017-x},
archivePrefix = {arXiv},
       eprint = {2212.07308},
 primaryClass = {astro-ph.SR},
       adsurl = {https://ui.adsabs.harvard.edu/abs/2023LRCA....9....2R}
}

@ARTICLE{Ge2024,
       author = {{Ge}, Hongwei and {Tout}, Christopher A. and {Chen}, Xuefei and {Wang}, Song and {Xiong}, Jianping and {Zhang}, Lifu and {Li}, Zhenwei and {Liu}, Qingzhong and {Han}, Zhanwen},
        title = "{Adiabatic Mass Loss in Binary Stars. V. Effects of Metallicity and Nonconservative Mass Transfer{\textemdash}Application in High Mass X-Ray Binaries}",
      journal = {\apj},
         year = 2024,
        month = nov,
       volume = {975},
       number = {2},
          eid = {254},
        pages = {254},
          doi = {10.3847/1538-4357/ad7ea6},
archivePrefix = {arXiv},
       eprint = {2408.16350},
 primaryClass = {astro-ph.SR},
       adsurl = {https://ui.adsabs.harvard.edu/abs/2024ApJ...975..254G}
}

@ARTICLE{webbink84,
       author = {{Webbink}, R.~F.},
        title = "{Double white dwarfs as progenitors of R Coronae Borealis stars and type I supernovae.}",
      journal = {\apj},
         year = 1984,
        month = feb,
       volume = {277},
        pages = {355-360},
          doi = {10.1086/161701},
       adsurl = {https://ui.adsabs.harvard.edu/abs/1984ApJ...277..355W}
}

@ARTICLE{ceres,
       author = {{Brahm}, Rafael and {Jord{\'a}n}, Andr{\'e}s and {Espinoza}, N{\'e}stor},
        title = "{CERES: A Set of Automated Routines for Echelle Spectra}",
      journal = {\pasp},
         year = 2017,
        month = mar,
       volume = {129},
       number = {973},
        pages = {034002},
          doi = {10.1088/1538-3873/aa5455},
archivePrefix = {arXiv},
       eprint = {1609.02279},
 primaryClass = {astro-ph.IM},
       adsurl = {https://ui.adsabs.harvard.edu/abs/2017PASP..129c4002B}
}

@ARTICLE{marcadon26,
       author = {{Marcadon}, F. and {Moharana}, A. and {Pawar}, T.~B. and {Pawar}, G. and {He{\l}miniak}, K.~G. and {Marques}, J.~P. and {Konacki}, M.},
        title = "{RX Gru: A short-period pre-main-sequence eclipsing binary with a distant circumbinary companion}",
      journal = {\aap},
         year = 2026,
        month = jul,
       volume = {710},
          eid = {A394},
        pages = {A394},
          doi = {10.1051/0004-6361/202557336},
archivePrefix = {arXiv},
       eprint = {2509.17011},
 primaryClass = {astro-ph.SR},
       adsurl = {https://ui.adsabs.harvard.edu/abs/2026A&A...710A.394M}
}

@dataset{wright19,
       author = {{Wright}, Edward L. and {Eisenhardt}, Peter R.~M. and {Mainzer}, Amy K. and {Ressler}, Michael E. and {Cutri}, Roc M. and {Jarrett}, Thomas and {Kirkpatrick}, J. Davy and {Padgett}, Deborah and {McMillan}, Robert S. and {Skrutskie}, Michael and {Stanford}, S.~A. and {Cohen}, Martin and {Walker}, Russell G. and {Mather}, John C. and {Leisawitz}, David and {Gautier}, III, Thomas N. and {McLean}, Ian and {Benford}, Dominic and {Lonsdale}, Carol J. and {Blain}, Andrew and {Mendez}, Bryan and {Irace}, William R. and {Duval}, Valerie and {Liu}, Fengchuan and {Royer}, Don and {Heinrichsen}, Ingolf and {Howard}, Joan and {Shannon}, Mark and {Kendall}, Martha and {Walsh}, Amy L. and {Larsen}, Mark and {Cardon}, Joel G. and {Schick}, Scott and {Schwalm}, Mark and {Abid}, Mohamed and {Fabinsky}, Beth and {Naes}, Larry and {Tsai}, ChaoWei},
        title = "{AllWISE Source Catalog}",
 howpublished = {NASA IPAC DataSet, IRSA1},
         year = 2019,
        month = jan,
          doi = {10.26131/IRSA1},
       adsurl = {https://ui.adsabs.harvard.edu/abs/2019ipac.data...I1W}
}

@ARTICLE{tandon17,
       author = {{Tandon}, S.~N. and {Subramaniam}, Annapurni and {Girish}, V. and {Postma}, J. and {Sankarasubramanian}, K. and {Sriram}, S. and {Stalin}, C.~S. and {Mondal}, C. and {Sahu}, S. and {Joseph}, P. and {Hutchings}, J. and {Ghosh}, S.~K. and {Barve}, I.~V. and {George}, K. and {Kamath}, P.~U. and {Kathiravan}, S. and {Kumar}, A. and {Lancelot}, J.~P. and {Leahy}, D. and {Mahesh}, P.~K. and {Mohan}, R. and {Nagabhushana}, S. and {Pati}, A.~K. and {Kameswara Rao}, N. and {Sreedhar}, Y.~H. and {Sreekumar}, P.},
        title = "{In-orbit Calibrations of the Ultraviolet Imaging Telescope}",
      journal = {\aj},
         year = 2017,
        month = sep,
       volume = {154},
       number = {3},
          eid = {128},
        pages = {128},
          doi = {10.3847/1538-3881/aa8451},
archivePrefix = {arXiv},
       eprint = {1705.03715},
 primaryClass = {astro-ph.IM},
       adsurl = {https://ui.adsabs.harvard.edu/abs/2017AJ....154..128T}
}

%%%%%%%%%%%%%%%%%%%%%%%%%%%%%%%%%%%%%%%%%%%%%%%%%%%%%%%
\begin{appendix}
   \nolinenumbers

\twocolumn

\section{Observations and data reduction}\label{appendix:data}

\subsection{Low-resolution spectroscopy}\label{sec:gmos}

Low resolution spectroscopy was carried out with the Gemini Multi-Object Spectrograph \citep[GMOS,][]{hook04} mounted at the Gemini South Telescope in Cerro Pach\'on, Chile. GMOS is equipped with a red-sensitive Hamamatsu detector with a pixel scale of 0.160 arcsec pixel$^{-1}$ for 2$\times$2 binning observations. 

Three spectra were taken during 2024 (under Gemini program GS-2024A-Q-412) with a 1\arcsec\,slit and both the B600 (two spectra) and B480 (one spectrum) gratings for resolving powers of $\sim$800 and $\sim$750, respectively. During 2025, 13 spectra were taken (under Gemini program GS-2025A-Q-311), 11 with the B600 grating, and two with B480, both with the same slit as before. In all cases the exposure time was fixed at 600 s, and the grating was centered at 520 nm. Heliocentric Julian dates (HJD) for all observations are given in Table \ref{tab:velocities}. In total, the timespan of these observations was longer than a year.

All spectra were bias subtracted, flat-fielded, and wavelength calibrated with DRAGONS \citep{dragons}, while the extraction, including background subtraction was handled with IRAF/apall.

\subsection{High-resolution spectroscopy}\label{sec:ghost}

A single high-resolution spectrum of the target was taken with the Gemini High-resolution Optical SpecTrograph \citep[GHOST,][]{kalari24} on the night of April 28, 2025. GHOST was used in the standard resolution mode for a resolving power of 56\,000. Exposure times were 1800 s for the blue arm, and 2$\times$900 s in the red arm. Detectors were read-out using a 1$\times$2 binning. GHOST covers simultaneously the optical range from 347 nm to 1060 nm. The spectra were also pipeline reduced by DRAGONS \citep{placco24} and retrieved from the Gemini Archive. This spectrum has a signal to noise ratio (SNR) of $\sim$110 at 450 nm.

Additionally, 3$\times$900\,s exposures were taken with the Bochum Echelle Spectrograph for OCA \citep[BESO,][]{steiner06}, significantly modified from its initial specifications, and now mounted on the Janusz  Ka{\l}u\.zny 1.5 m telescope at the Rolf Chini Cerro Murphy Observatory in Chile\footnote{https://oca.camk.edu.pl/} on May 21, 2026. BESO provides a spectral resolution of 48\,000. BESO spectra were reduced with the CERES pipeline \citep{ceres} which was adapted to reduce BESO spectra.

\subsection{Radial velocity measurements}

RVs for all epochs, of both GMOS and GHOST spectra, were measured identifying individual lines in the spectra with IRAF/rvid; mostly of H and He I which dominate spectra at this temperature. Only the broad H$\alpha$ and H$\beta$ were excluded from the measurements. On average, velocities were measured from 22 absorption lines. The BESO spectrum had significantly lower SNR and therefore its velocity was measured via cross-correlation against the Doppler corrected GHOST spectrum. Epoch velocities can be seen in Table \ref{tab:velocities}.

\begin{table}[t]
\caption{\label{tab:velocities}Log of GMOS observations and their respective heliocentric velocities}
\centering
\begin{tabular}{lrr}
\hline
\hline
HJD & Radial velocity  & Unc. \\
(day) & (km s$^{-1}$) & (km s$^{-1}$)\\
\hline
2460410.58021 & 25.73 & 2.70\\
2460411.57608 & --5.84 & 3.33\\
2460469.45601 & 26.97 & 4.35\\
2460737.59256 & 8.96 & 2.46\\
2460751.49584 & 11.87 & 1.67\\
2460759.52084 & --20.92 & 2.59\\
2460764.62721 & --18.20 & 3.08\\
2460771.61490 & 7.90 & 3.11\\
2460773.53500 & 9.73 & 4.11\\
2460777.49218 & 33.32 & 3.32\\
2460780.51947 & 29.82 & 3.12\\
2460785.55287 & 12.25 & 2.22\\
2460791.55295 & --9.14 & 6.34\\
2460794.65682 & --20.63 & 8.19\\
2460798.48235 & --12.02 & 3.22\\
2460802.59683 & --8.84 & 2.35\\
\hline
\end{tabular}
\end{table}

\begin{table}[h]
\caption{\label{tab:target}Target properties}
\centering
\begin{tabular}{lrr}
\hline\hline
Parameter & Value &  Ref.\\
\hline
Gaia DR3 ID   &  5203319520096719360   &\tablefootmark{a}\\
RA (deg)  &   140.2000366049  &\tablefootmark{a}\\
Dec (deg) &    --77.2542358615 &\tablefootmark{a}\\
Parallax (mas) & 0.1138$\pm$0.0127 &\tablefootmark{a}\\
PM$_{\textrm {RA}}$ (mas year$^{-1}$) & --2.6651$\pm$0.0155  &\tablefootmark{a}\\
PM$_{\rm {Dec}}$    (mas year$^{-1}$) &   7.0414$\pm$0.0155  &\tablefootmark{a}\\
\hline
$M_1$ ($M_{\odot})$& 0.53 $\pm$ 0.02 &\tablefootmark{b}\\
Orbital period (days) & 34.04 $\pm$ 0.05 &\tablefootmark{c}\\
Eccentricity & 0.119 $\pm$ 0.04 & \tablefootmark{c}\\
Barycentric velocity (km s$^{-1}$) & 5.13 $\pm$ 0.69 &\tablefootmark{c}\\
$K_1$ (km s$^{-1}$) & 24.85 $\pm$ 1.24& \tablefootmark{c} \\
$f(m)$ ($M_{\odot}$) & 0.053 $\pm$ 0.008& \tablefootmark{c}\\
\hline
$T_{\rm eff}$ (K)& 18\,500 $\pm$ 700 & \tablefootmark{c}\\
$\log g$ (cm s$^{-2}$)&2.70 $\pm$ 0.25 & \tablefootmark{c}\\
\hline
\end{tabular}
\tablefoot{.\tablefoottext{a}{Gaia DR3}
\tablefoottext{b}{\citetalias{kumar24}}
\tablefoottext{c}{This work}
}
\end{table}

\section{Posterior distribution of orbital parameters}\label{appendix:corner}
   \begin{figure}[]
   \centering
   \includegraphics[width=0.5\textwidth]{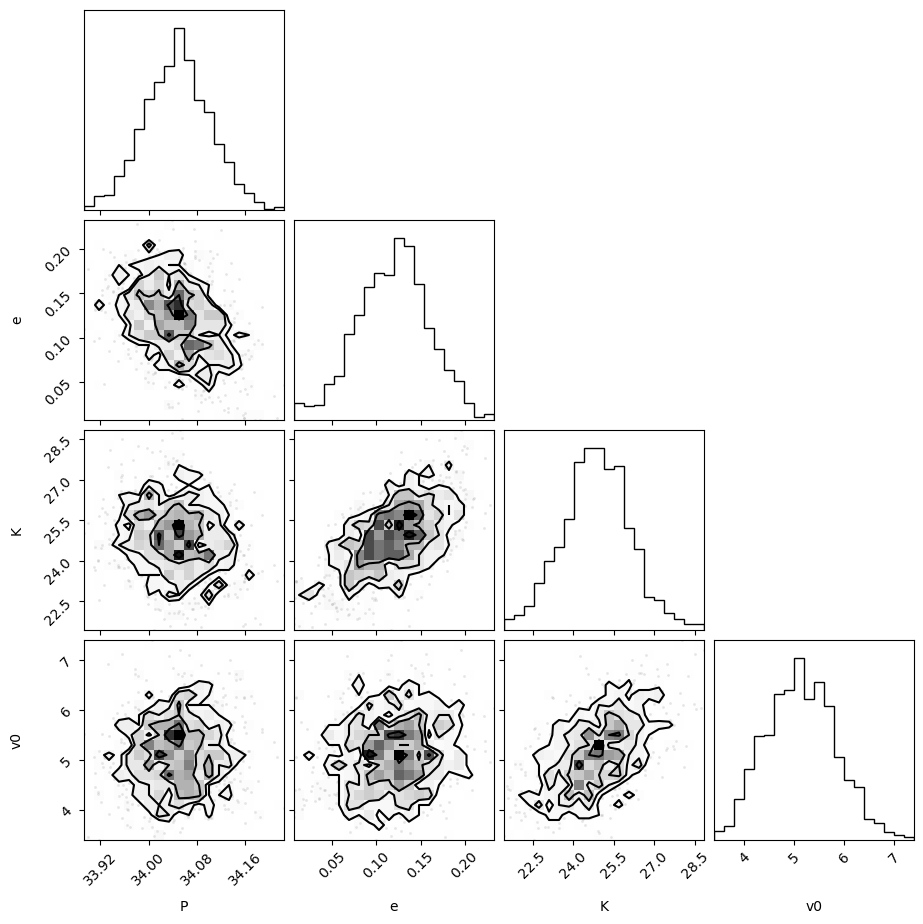}
      \caption{Posterior distribution of the eccentricity, RV semi-amplitude and mean velocity from \textsc{TheJoker}.}
         \label{fig:corner}
   \end{figure}

\section{Stellar parameter determination}\label{appendix:stellar}

We constrained $T_{\rm eff}$ and $\log g$ following the general methodology of \citetalias{kumar24}. The Balmer-line wings were first used to define a locus
of acceptable surface gravities, $\log g(T_{\rm eff})$, across the full model grid. The available Si\,\textsc{ii} and Si\,\textsc{iii} profiles were subsequently evaluated along this locus as consistency and systematic-error diagnostics. We did not, however, use these lines to derive a formal Si
ionisation-balance solution, nor did we combine the Si and Balmer-line
residuals into a single $\chi^2$ statistic. This choice was motivated by the
strong dependence of the inferred Si abundance on the ion and multiplet
considered and by the limited differential
temperature sensitivity of the two acceptable profiles, which prevented the Si lines from defining a robust
independent statistical minimum.

The gravity constraint was obtained primarily from H$\beta$.  Synthetic profiles were calculated over a broad interval and compared with the
observed wings after excluding the line core.  To quantify the sensitivity to continuum placement and to the adopted wing mask, we fitted a
multiplicative linear correction simultaneously with each synthetic profile. We tested inner exclusions of $|\Delta\lambda|=3$, 5, 7, 10, and 12~\AA\ and outer limits of $|\Delta\lambda|=25$ and 35~\AA.  The masks retaining the
gravity-sensitive inner wings ($|\Delta\lambda|=3$--7~\AA) consistently
favoured the $\log g=2.75$ grid model at $T_{\rm eff}=18\,000$~K.  At
$19\,000$~K the preferred models lay between $\log g=2.75$ and 3.00, depending on the exact mask.  Removing the central 10--12~\AA\ made the gravity constraint nearly degenerate because most of the informative inner wings had then been discarded; those masks were used only as robustness tests. H$\alpha$ was analysed in the same manner as an independent consistency check, but was not assigned equal weight because its local normalisation and core profile are less reliable in the GHOST spectrum. We also calculated
profiles for the \citetalias{kumar24} spectroscopic pair, $T_{\rm eff}=17\,000$ K and $\log g=2.37$. This model produces systematically narrower H$\beta$ wings than observed and gives a mean squared residual 1.54 times larger than our
adopted solution over $5<|\Delta\lambda|<35$~\AA. At
$T_{\rm eff}=17\,000$ K, the Balmer locus instead favours
$\log g\simeq2.7$. In all cases, the Balmer cores were excluded from the parameter determination.

We next calculated silicon abundance grids at each available
$(T_{\rm eff},\log g)$ node, using the corrected BSTAR2006/SYNSPEC abundance
conversion $A_{\rm native}=-10^{[\mathrm{Si/H}]}$.  The syntheses covered
6335--6385~\AA\ for Si~{\sc ii} and 4545--4585~\AA\ for Si~{\sc iii}.

We repeated the \citetalias{kumar24} ionisation-balance test using the same five nominal
diagnostics: Si~{\sc ii} $\lambda\lambda6347,6371$ and Si~{\sc iii} 
$\lambda\lambda4552,4567,4575$. The last transition corresponds to
4574.757~\AA\ in our adopted VALD list. Only Si~{\sc ii} $\lambda6371$ and Si~{\sc iii} $\lambda4575$ satisfy our predefined GHOST profile-quality criteria; Si~{\sc ii} $\lambda6347$ and Si~{\sc iii} $\lambda\lambda4552,4567$ were rejected because of profile and/or depth mismatches. At the \citetalias{kumar24} atmospheric pair, the two accepted lines differ by 1.25 dex in the inferred silicon abundance and do not reproduce the \citetalias{kumar24} ionisation-balance crossing. Moreover, these two profiles alone do not discriminate robustly between 17\,000 and 18\,500 K.
Silicon was therefore retained as a consistency and systematic-error
diagnostic rather than as an independent atmospheric-parameter constraint.

An additional examination of silicon transitions outside the two Si ii lines used by \citetalias{kumar24} reveals the same ionisation-stage discrepancy. Near the adopted stellar parameters, Si ii $\lambda\lambda5041,5056$ yield abundances approximately 1.4--1.7 dex lower than those inferred from Si iii $\lambda\lambda4568,4575$. This discrepancy persists
after examining the atomic data, possible blends, local continuum placement,
equivalent-width measurements, and the adopted microturbulence. Although
increasing $T_{\rm eff}$ reduces the disagreement, models with
$T_{\rm eff}=21\,000$--$22\,000$\,K predict a Si\,\textsc{iv}
$\lambda 4116$ feature substantially stronger than observed. We therefore use
the silicon lines only as qualitative consistency checks and as indicators of
systematic uncertainty, rather than including them as formal constraints in
the stellar-parameter determination.

The Balmer-line analysis identifies an allowed region in the
$T_{\rm eff}$--$\log g$ plane rather than a unique statistical minimum. The
Balmer wings provide the primary constraints on the atmospheric parameters,
while the usable silicon profiles serve only as qualitative consistency
checks. In particular, the absence of the Si\,\textsc{iv} $\lambda 4116$
feature at the strength predicted by the hotter models disfavors solutions
with $T_{\rm eff}\gtrsim 21\,000$\,K. Taking these diagnostics into account,
without including the silicon lines as formal contributions to the fit, we
adopted the representative solution

\begin{equation}
    T_{\rm eff}=18\,500\pm700\,{\rm K},\qquad
    \log g=2.70\pm0.25 .
\end{equation}

The adopted gravity lies close to the well-constrained $\log g=2.75$ grid
node. Its quoted uncertainty encompasses the variations associated with
temperature, the choice of Balmer-wing masks, local continuum placement, and
the consistency check provided by H$\alpha$. The temperature uncertainty
covers the range of models providing acceptable Balmer-wing fits and also
reflects the limited temperature constraint supplied by the silicon
profiles. These uncertainties should therefore be regarded as systematic
ranges that account for the diagnostic sensitivity and finite grid
resolution, rather than as formal confidence intervals derived from the
curvature of a single $\chi^2$ surface. The silicon abundance reported in Appendix \ref{appendix:abundances}
is an adopted profile-based estimate and should not be interpreted as the
result of a successful Si\,\textsc{ii}/Si\,\textsc{iii} ionisation
equilibrium.

\begin{figure*}
    \centering
    \includegraphics[width=0.75\textwidth]{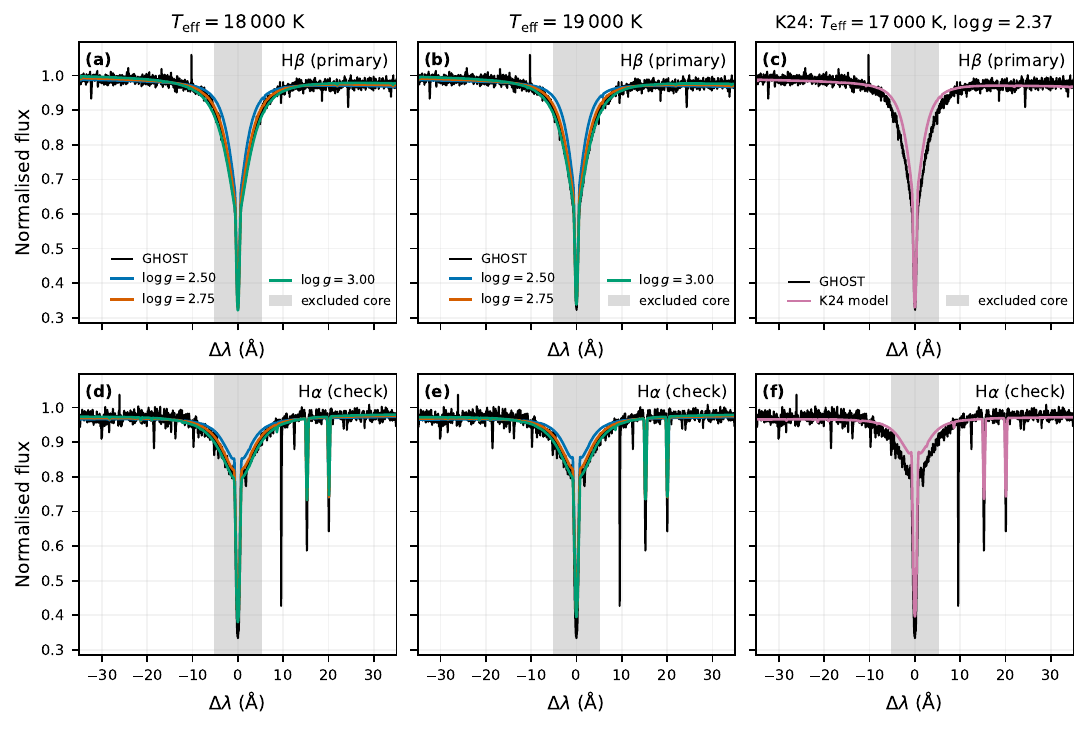}
    \caption{Comparison of the observed GHOST Balmer-line profiles with synthetic spectra calculated for different surface gravities. The upper panels (a)--(c) show H$\beta$, which provides the primary gravity constraint, whereas the lower panels (d)--(f) show H$\alpha$, used as an independent consistency check. The first and second columns show models with $T_{\rm eff}=18\,000$ and $19\,000$~K, respectively, for $\log g=2.50$, 2.75, and 3.00. The third column shows the K24 spectroscopic parameters, $T_{\rm eff}=17\,000$~K and $\log g=2.37$. The observed spectrum is shown in black; the synthetic profiles for $\log g=2.50$, 2.75, and 3.00 are shown in blue, orange, and green, respectively, and the K24 profile is shown in magenta. The shaded regions indicate the excluded line cores; only the Balmer-line wings were considered in the atmospheric-parameter analysis.}
      \label{fig:balmer_diagnostics}
  \end{figure*}
  
\section{Chemical abundances}\label{appendix:abundances}

We determined abundances by direct comparison of the GHOST spectrum with synthetic spectra based on NLTE BSTAR2006 model atmospheres,
calculated with \textsc{synspec}, and convolved with \textsc{rotin3}.  The atmospheric parameters and projected rotational velocity were held fixed at $T_{\rm eff}=18\,500\pm700$~K, $\log g=2.70\pm0.25$, and $v\sin i\lesssim5$~km\,s$^{-1}$; the BSTAR2006 microturbulent velocity was
fixed at $\xi=2$~km\,s$^{-1}$.   For each accepted transition, the abundance was varied over the adopted grid, and the observed and synthetic line profiles were compared using a consistent local treatment.  Lines affected by blends, continuum defects, detector artefacts, wavelength offsets, weak abundance sensitivity, or synthesis-window boundaries were rejected.

The adopted atmosphere lies between four BSTAR2006 grid points.  We therefore evaluated each abundance diagnostic at $(T_{\rm eff},\log g)=(18000,2.50)$, $(18000,2.75)$, $(19000,2.50)$, and $(19000,2.75)$, and combined the line results with bilinear weights of 0.10, 0.40, 0.10, and 0.40, respectively. This procedure interpolates the results of independent line syntheses rather than the atmospheric structures.  For the six selected elements, we examined each candidate spectral line individually and included in the abundance analysis only those with reliable synthetic profiles at all four atmospheric-grid nodes used to interpolate the adopted stellar parameters.
For C, N, O, Al, and Fe, the tabulated $\sigma_{\rm L}$ values are the line-to-line standard deviations of the retained transitions.
For Si, the quoted uncertainty is a local profile-fitting precision; the larger ion- and multiplet-dependent discrepancy is retained as an explicit systematic limitation.  The quoted uncertainties are not formal confidence intervals or total errors, and larger atmospheric, multiplet, or ionisation-stage differences are treated as systematic diagnostics.

The resulting abundances are listed in  Table~\ref{tab:abundances}.  From four accepted optical Fe features we obtain
$[\mathrm{Fe/H}]=-0.85\pm0.17$, where the uncertainty is the line-to-line
dispersion.  Relative to the iron abundance, we obtain $[\mathrm{C/Fe}]=-0.07$, $[\mathrm{N/Fe}]=+0.61$, $[\mathrm{O/Fe}]=+0.53$, $[\mathrm{Al/Fe}]=+0.19$, and $[\mathrm{Si/Fe}]=+0.65$.

The homogeneous maximum-line analysis retained 8 C~{\sc ii}, 12 N~{\sc ii}, 16 O~{\sc ii}, 4 Al~{\sc iii}, and 4 Fe~{\sc ii}/Fe~{\sc iii} lines at all four atmospheric nodes. For Si, $N=2$ denotes the two profile diagnostics supporting the adopted value; additional lines were
inspected but not averaged because they separate into incompatible ionisation or multiplet groups.

We required at least two aligned, abundance-sensitive transitions to give abundances consistent within 0.30 dex before identifying an $s$-process detection.  None of Sr, Y, Zr, or Ba met this criterion, and none of the individual lines is adequate for a reliable abundance estimate.
Sr~{\sc ii} 4215.52~\AA\ shows a weak aligned feature, but its isolated
profile minimum would imply an extreme abundance and is not corroborated by
Sr~{\sc ii} 4077.71~\AA.  The Y~{\sc ii} diagnostics are compromised by
blue-region noise, wavelength offsets, or blends.  The available
Zr~{\sc iii} lines arise from highly excited lower levels and mostly show
negligible abundance sensitivity; responsive profiles fail the alignment or
morphology tests.  The Ba~{\sc ii} resonance lines are non-detections and
Ba~{\sc ii} 6496.90~\AA\ is too weak for a secure measurement.

At $T_{\rm eff}\simeq18500$~K, the commonly used Sr~{\sc ii}, Y~{\sc ii}, and Ba~{\sc ii} optical lines also sample minority ionisation stages.  The lack of useful lines from the dominant stages and the use of
trace LTE heavy species introduce an additional, poorly quantified ionisation/NLTE systematic.  Exploratory profile bounds were extremely high and model dependent and therefore do not constitute astrophysically useful upper limits.  We report no abundance or upper limit for these
elements: the present data neither exclude ordinary $s$-process enrichment
nor decide whether third dredge-up occurred.

Using the solar scale of \citet{AsplundEtAl2009}, the CNO abundances correspond to $\log\epsilon(\mathrm{C})=7.52$, $\log\epsilon(\mathrm{N})=7.59$, and
$\log\epsilon(\mathrm{O})=8.37$.  Thus, $\log(\mathrm{N/O})=-0.78$, $\log(\mathrm{C/O})=-0.85$, and $\mathrm{C/O}\simeq0.14$ by number.  The photosphere is therefore oxygen-rich
and nitrogen is enhanced relative to the overall iron deficiency.  The total CNO abundance is $\log\epsilon(\mathrm{C+N+O})\simeq8.49$, or approximately
$[\mathrm{(C+N+O)/H}]=-0.43$ on the same solar scale.

The C~{\sc ii} region contains additional unmatched structure, and the
oxygen abundance is comparatively sensitive to the adopted atmospheric
parameters.  Individual Si~{\sc ii} and Si~{\sc iii}
re-minimisations do not provide a perfectly consistent ionisation balance.  Although the adopted combined abundance, $[\mathrm{Si/H}]=-0.20\pm0.12$, is very similar to \citetalias{kumar24}, agreement in the mean abundance does not imply agreement in the line-by-line Si~{\sc ii} and Si~{\sc iii} equilibrium. The quoted uncertainty represents the internal precision of the accepted profile fits and does not include the larger ion-, multiplet-, atomic-data-, or NLTE-dependent systematics.

\begin{figure}
    \centering
    \includegraphics[width=\columnwidth]{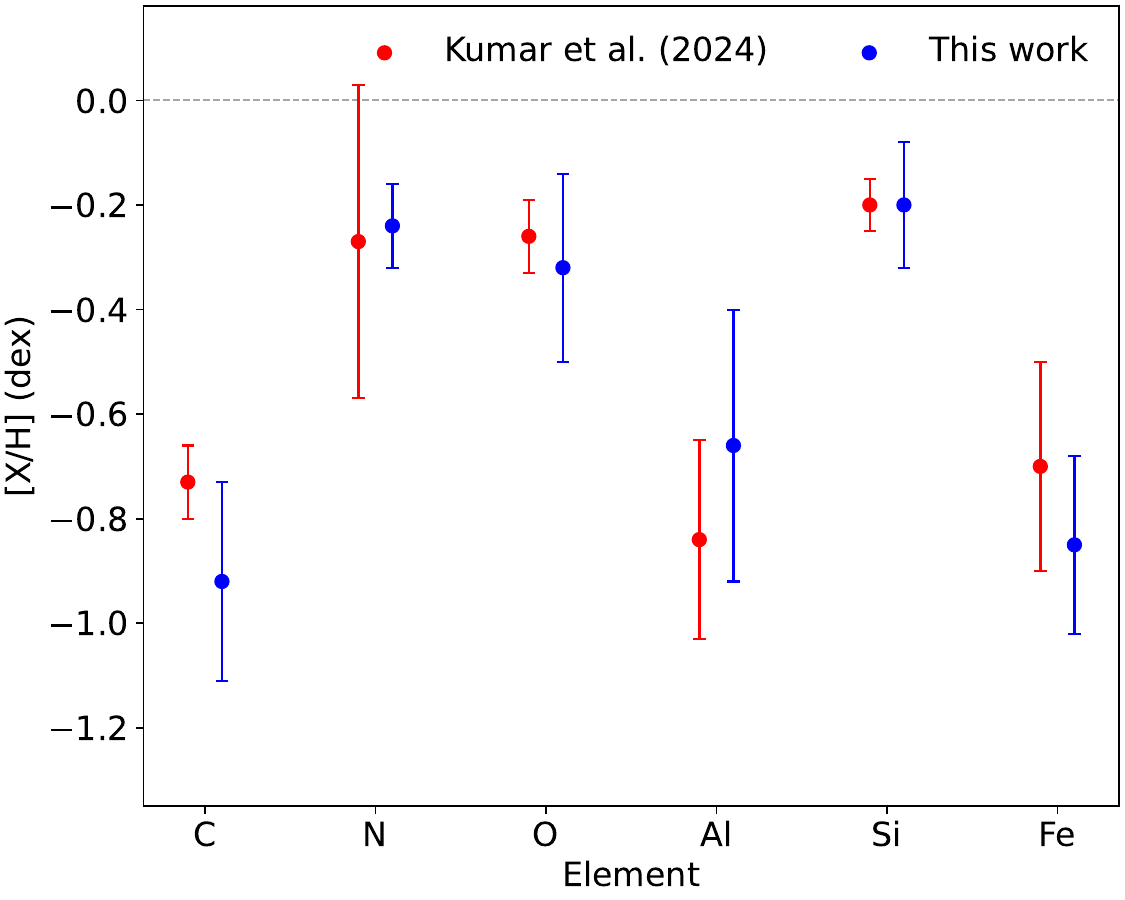}
    \caption{Comparison between the abundances obtained in this work and
those reported by \citetalias{kumar24}. Turquoise circles represent the abundances from
\citetalias{kumar24}, while deep-pink squares show the results of the present analysis. The error bars correspond to the uncertainties quoted
in the respective analyses. The horizontal dashed line indicates the solar abundance ratio, $[\mathrm{X/H}]=0$.}
    \label{fig:abundances}
  \end{figure}

\section{Period-eccentricity}\label{appendix:period}

\begin{figure}[H]
   \centering
   \includegraphics[width=\columnwidth]{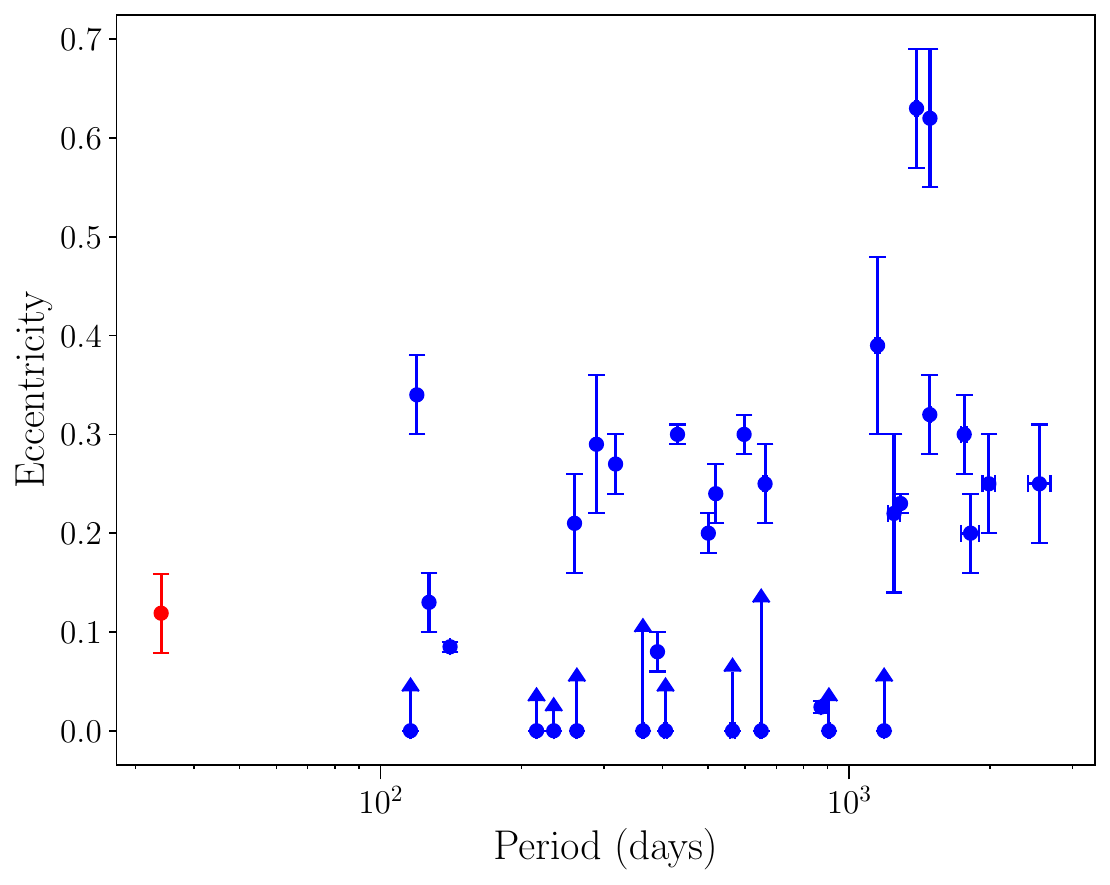}
      \caption{Period-eccentricity plot. Blue symbols indicate 33 post- AGB in binary systems studied by \citet{oomen18}, while the red symbol shows the pAGB studied here.}
         \label{fig:periods}
   \end{figure}

\section{\textsc{compas} simulations}\label{appendix:compas}

\textsc{compas} was used under its \texttt{BSE} (binary stellar evolution) mode, with the following changes to default binary evolution and physics parameters:
\begin{itemize}
    \item Maximum evolution time is set at $13.5$ Gyr, covering the cluster's age estimates between 10 and 13 Gyr \citep[e.g.,][]{FuenteMarcos2015}.
    \item We set the metallicity to $Z=0.0026$, based on the [Fe/H] value presented in \cite{monaco18}, and assuming a simple [Fe/H]$\approx\log_{10}\left(Z/Z_\odot\right)$ relation, using $Z_\odot=0.02$.
    \item We explore four different values for the common envelope efficiency parameter, $\alpha_{\rm CE}$: 0.1, 1.0, 3.0, and 5.0. The lowest value follows typical results from post-common envelope studies \citep[e.g.,][]{Zorotovic2010}, whereas higher values are motivated by uncertainties surrounding the treatment of the common envelope phase and possible additional sources of energy contributing to envelope removal \citep[see e.g.,][]{Ropke2023}.
    \item Critical mass ratios that determine the onset of dynamically unstable mass transfer (common envelope) are drawn from \citet[][and references therein]{Ge2024}, which are based on stellar adiabatic mass-loss sequences.
    \item Angular momentum is set to be lost from the position of the accreting star (usually referred to as isotropic re-emission), with the accretion rate (similarly, binary mass loss rate) limited by the thermal timescale \citep[see e.g., Section 4.2.3 in][and references therein]{riley22}.
\end{itemize}

\subsection{Grid of binaries}

Since we are interested in generating possible evolutionary paths for the pAGB system in E3, we limit ourselves to evolving a custom grid of binaries defined by a comparatively narrow range of primary mass, mass ratio, and orbital period values than those of typical BPS studies. The grid parameters are summarised in Table~\ref{tab:grid} and further explained below.
 
\begin{table*}
\caption{Binary grid parameters.\label{tab:grid}}
\centering
\begin{tabular}{llcc}
\hline
\hline
Parameter & Range & Sampling & Number of values\\
\hline
Primary mass, $M_1$ & 0.5--2.0 $\msun$ & Linear, $\Delta M_1 = 0.012\,\msun$ & 126 \\
Mass ratio, $q = M_2/M_1$ & 0.01--1.0 & Linear, $\Delta q = 0.04125$ & 25 (before cut) \\
Orbital period, $P$ & 0.1--1000 d & Logarithmic, $\Delta\log_{10}P \approx 0.0201$ & 200 \\
Eccentricity, $e$ & 0 & Fixed & 1 \\
\hline
\end{tabular}
\tablefoot{After the cut $M_2 \geq 0.08\,\msun$, 2,895 of the 3,150 $(M_1, q)$ combinations remain, giving 2\,895 $\times$ 200 = 579\,000 models in total.}
\end{table*}
 
The primary mass is sampled at 126 linearly spaced values between 0.5 and 2.0\,$\msun$. At each primary mass, the mass ratio $q$ is sampled at 25 linearly spaced values between 0.01 and 1.0, defining the stellar companion's mass as $M_2 = q\times M_1$. Systems resulting in $M_2 < 0.08\,\msun$ are excluded, to avoid low-mass extrapolation in \textsc{compas}. Each mass pair is then evaluated at 200 initial orbital periods, logarithmically spaced between 0.1 and 1000\,d. All orbits are initially set as circular ($e = 0$) for simplicity, and no eccentricity evolution is enforced.
 
The resulting grid contains $2\,895 \times 200 = 579\,000$ binary models, where $2\,895$ corresponds to the total number of mass pairs after removing $M_2 < 0.08\,\msun$.

\subsection{Refining the grid}
 
 \begin{figure}[t!]
   %\centering
   \includegraphics[width=0.5\textwidth]{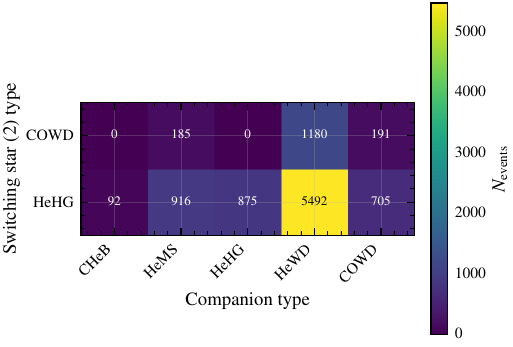}
      \caption{Annotated heatmap of the COMPAS simulations for the $\alpha_{\rm CE}=3.0$ set. The x-axis indicates the pAGB companion type, while the y-axis indicates the evolutionary path of the pAGB system. The abbreviations denote core helium burning (CHeB), helium main sequence (HeMS), helium Hertzsprung gap (HeHG), helium white dwarf (HeWD), and carbon–oxygen white dwarf (COWD). For details about these stellar types, see \citet{riley22}.}
         \label{fig:types}
   \end{figure}
   
The results of the evolved grid cover a wide range of configurations. However, while we are interested in pAGB systems, there is no such definition within the \textsc{compas} stellar types and, therefore, we look for the most similar definition available. Under this approach, we focus on stars evolving from either the early AGB or thermally pulsing (TP) AGB into a non-AGB phase (that is, eAGB to TP-AGB transitions are excluded). Next, we impose additional restrictions to the orbital period, namely $10 \leq P_{\rm orb} \leq 200\,{\rm d}$. This results in selecting systems that match the orbital period measurement for the pAGB in E3, as well as shorter and larger values to account for the many uncertainties derived from BPS.

\end{appendix}
\end{document}